\documentclass[prl,twocolumn,letterpaper, superscriptaddress,longbibliography]{revtex4-1}
\usepackage{graphicx}
\usepackage{dcolumn}
\usepackage{bm}
\usepackage[usenames,dvipsnames]{color} 
\usepackage{tabularx}
\usepackage{array}
\usepackage{amsmath}
\usepackage{pbox}
\usepackage{stmaryrd}  
\usepackage{floatrow}
\usepackage{sidecap}
\usepackage{gensymb}
\usepackage{soul}
\usepackage{physics}

\NewDocumentCommand{\stau}{}{%
  \mathord{\text{\resizebox{!}{1ex}{\usefont{U}{psy}{m}{n}\symbol{"74}}}}%
}

\begin{document}

\title{Tuning Magneto-Optical Zero-Reflection via Dual-Channel Hybrid Magnonics}

\author{Andrew Christy}
\thanks{These authors contributed equally.}
\affiliation{Department of Physics and Astronomy, University of North Carolina at Chapel Hill, Chapel Hill, NC 27599, USA}
\affiliation{Department of Chemistry, University of North Carolina at Chapel Hill, Chapel Hill, NC 27599, USA}

\author{Yujie Zhu}
\thanks{These authors contributed equally.}
\affiliation{Department of Materials Science and Engineering, University of Wisconsin-Madison, Madison, Wisconsin, 53706, USA}

\author{Yi Li}
\affiliation{Materials Science Division, Argonne National Laboratory, Argonne, IL 60439, USA}

\author{Yuzan Xiong}
\affiliation{Department of Physics and Astronomy, University of North Carolina at Chapel Hill, Chapel Hill, NC 27599, USA}
\affiliation{Department of Electrical and Computer Engineering, North Carolina A$\&$T State University, Greensboro, NC 27411, USA}

\author{Tao Qu}
\affiliation{Department of Physics and Astronomy, University of North Carolina at Chapel Hill, Chapel Hill, NC 27599, USA}

\author{Frank Tsui}
\affiliation{Department of Physics and Astronomy, University of North Carolina at Chapel Hill, Chapel Hill, NC 27599, USA}

\author{James F. Cahoon}
\affiliation{Department of Chemistry, University of North Carolina at Chapel Hill, Chapel Hill, NC 27599, USA}

\author{Binbin Yang}
\affiliation{Department of Electrical and Computer Engineering, North Carolina A$\&$T State University, Greensboro, NC 27411, USA}

\author{Jia-Mian Hu}
\thanks{Email: jhu238@wisc.edu}
\affiliation{Department of Materials Science and Engineering, University of Wisconsin-Madison, Madison, Wisconsin, 53706, USA}

\author{Wei Zhang}
\thanks{Email: zhwei@unc.edu}
\affiliation{Department of Physics and Astronomy, University of North Carolina at Chapel Hill, Chapel Hill, NC 27599, USA}

\date{\today}

\begin{abstract}

Multi-channel coupling in hybrid systems makes an attractive testbed not only because of the distinct advantages entailed in each constituent mode, but also the opportunity to leverage interference among the various excitation pathways. Here, via combined analytical calculation and experiment, we demonstrate that the phase of the magnetization precession at the interface of a coupled yttrium iron garnet(YIG)/permalloy(Py) bilayer is collectively controlled by the microwave photon field torque and the interlayer exchange torque, manifesting a coherent, dual-channel excitation scheme that effectively tunes the magneto-optic spectrum. The different torque contributions vary with frequency, external bias field, and types of interlayer coupling between YIG and Py, which further results in destructive or constructive interferences between the two excitation channels, and hence, selective suppression or amplification of the hybridized magnon modes.

\end{abstract}

\maketitle

\setlength{\belowdisplayskip}{0pt} \setlength{\belowdisplayshortskip}{0pt}
\setlength{\abovedisplayskip}{0pt} \setlength{\abovedisplayshortskip}{0pt}

\section{I. Introduction}

Hybrid magnonic systems have emerged as versatile modular components for quantum signal transduction and sensing applications owing to their capability of connecting distinct quantum platforms \cite{awschalom2021quantum}. Hybridization of magnons with phonons, microwave photons, qubits, single spins, and optical light have been demonstrated, with manifestations of characteristic magnon-coupled spectra, such as level repulsion/attraction \cite{goryachev2014high,zhang2014strongly,tabuchi2014hybridizing,wang2020dissipative,harder2018level}, magnetically-induced transparency(MIT) \cite{xiong2020probing,xiong2022tunable,xiong2024phase}, super/ultrastrong couplings \cite{inman2022hybrid}, pump-induced nonlinearity \cite{rao2023unveiling,yang2025control,xiong2024magnon,qu2025pump}, zero-reflection(ZR) \cite{qian2023non}, and spectrum singularities \cite{yang2020unconventional,zhang2019experimental,xiong2025photon}, bestowing emerging quantum engineering functionalities \cite{lachance2019hybrid,li2020hybrid,yuan2022quantum,flebus20242024,chumak2022advances}.

\begin{figure}[htb]
 \centering
 \includegraphics[width=3.4 in]{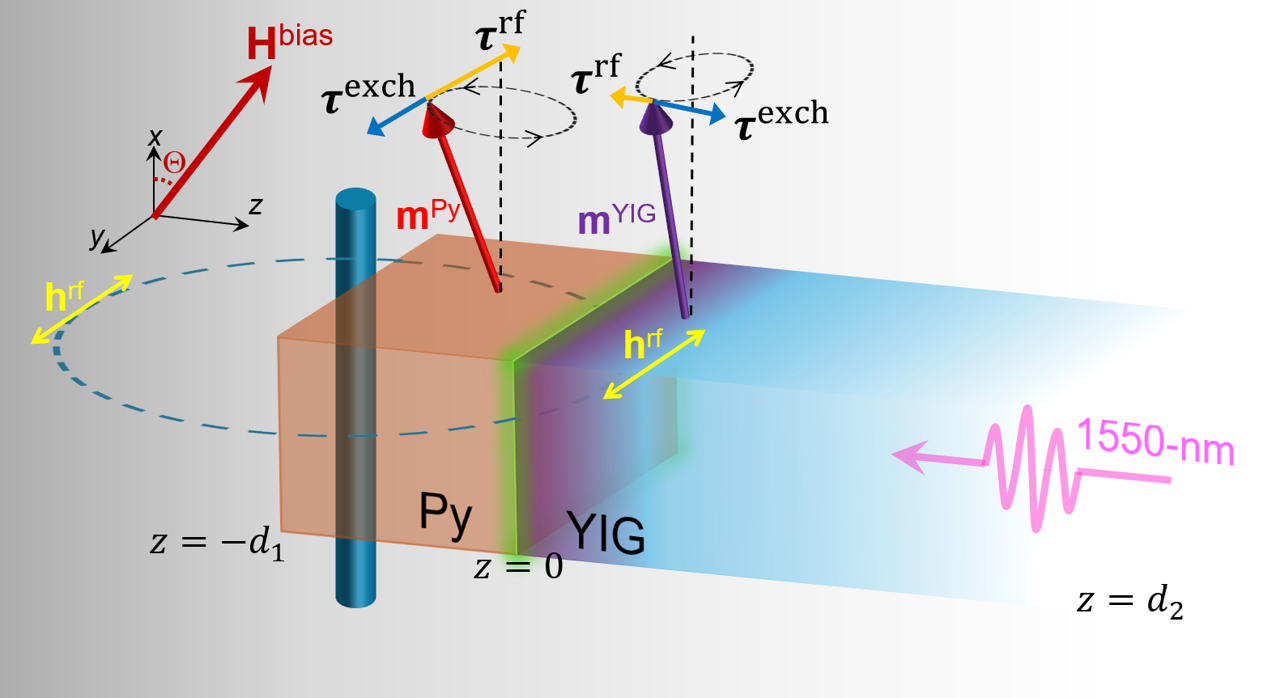}
 \caption{Schematic illustration of a coupled Py/YIG bilayer subject to a dual-channel excitation: the MW field ${\rm \bf h}_{\rm rf}$ excites the Py Kittel mode and the YIG PSSW modes (photon channel) via the Zeeman torque, $\stau^{\rm rf}$, and concurrently, the excited Py resonance dynamically couples to the YIG PSSW modes (magnon channel) via the exchange torque $\stau^{\rm exch}$ due to interlayer magnon-magnon coupling. ${\rm \bf H}^{\rm bias}$ is the external bias magnetic field that can rotate out of plane, at $\Theta$ with respect to the $x$ axis.}  
 \label{fig1}
\end{figure}

Most studies to date, however, only focused on a single coupling channel, e.g., photon-magnon \cite{goryachev2014high,zhang2014strongly,tabuchi2014hybridizing}, phonon-magnon \cite{liao2024nonreciprocal,li2021advances,matsumoto2024magnon}, or magnon-magnon \cite{li2020coherent,liensberger2019exchange,qin2018exchange,xiong2020probing} couplings, respectively. Extended tunabilities and/or advanced functionalities can further arise by allowing different coupling channels to combine and interfere, for example, nonreciprocal transmissions arise from engineering coherent and dissipative coupling channels in photon-magnon systems \cite{zhang2020broadband,wang2019nonreciprocity}, and, effective damping modulation from combining interfacial exchange and spin pumping channels in magnon-magnon systems \cite{li2020coherent,liu2024strong,subedi2025engineering,fan2025dynamically}.

\begin{figure*}[htb]
 \centering
 \includegraphics[width=6.9 in]{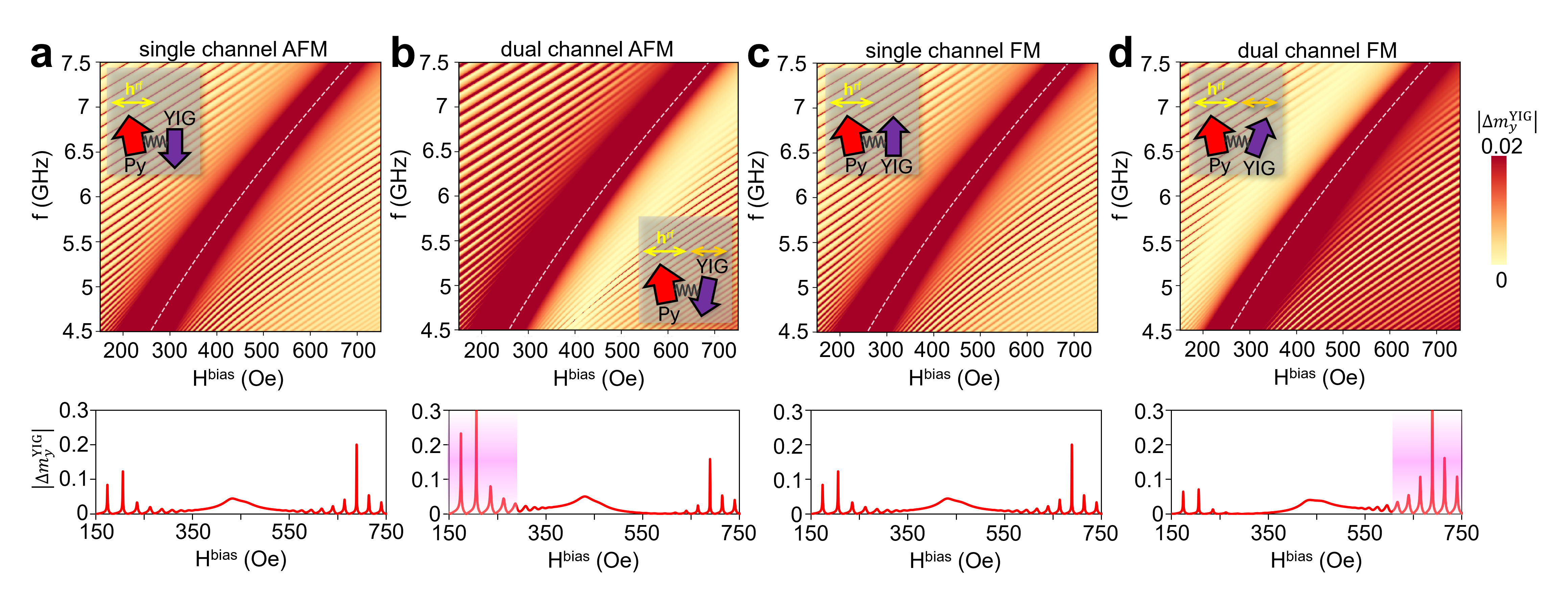}
 \caption{(a,b) The calculated $f - H$ dispersions of the YIG excitation (represented by the magnetization oscillation amplitude $\Delta {m}_y^{\rm YIG}$) for (a) single-channel (magnon only) and (b) dual-channel (magnon and photon) schemes, respectively, under an AFM-type interlayer coupling between YIG and Py. (c,d) The same calculation for (c) single-channel and (d) dual-channel schemes, but under an FM-type interlayer coupling. The lower panel shows example 1D signal-line trace at 6-GHz for each case. \textcolor{black}{The (pink) shadowed area highlights the amplified regime.} }  
 \label{fig2}
\end{figure*}

Such channel interference can, in principle, originate from disparate hybridization schemes. For example, a finite-wavelength magnon modes in low damping Y$_3$Fe$_5$O$_{12}$(YIG) can be subject to concurrent excitations from a non-uniform microwave(MW) field given by a stripline (photon channel) and an interfacial exchange field (magnon channel) via coupling to an adjacent Permalloy(Py) layer, as depicted in Fig.\ref{fig1}. Depending on the frequency, bias magnetic field, types of the interlayer coupling between YIG and Py, the co-operational Zeeman ($\stau^{\rm rf}$) and exchange ($\stau^{\rm exch}$) torques can lead to destructive or constructive interferences, hence, resulting in either suppressed or amplified effective excitations of the finite-wavelength modes of YIG. \textcolor{black}{Although the fundamentals in magnetization dynamics excited by the Zeeman and exchange torques are well established \cite{reid2010optical,kirilyuk2010ultrafast,kirilyuk2013laser,lalieu2017absorption}, such a coherent interference in manipulating distinct magnon modes and mode profiles has not been fully understood.} 

\section{II. Theoretical Model}

We consider a model picture in Fig.\ref{fig1} where a thin layer of Py is exchange-coupled to a relatively thick layer of YIG, stacked along $+z$. The interface between the Py and YIG is defined as $z=0$. The Py has a thickness $d_1$ (15 nm), and positioned between $z=-d_1$ and 0. The YIG has a thickness $d_2$ (3 $\mu$m), and positioned between 0 and $z=d_2$. The bias magnetic field, ${\rm \bf H}^{\rm bias}$, is initially applied along $+x$ ($\Theta = 0^\circ$) but can be rotated within the $xz$ plane. The MW field ${\rm \bf h}^{\rm rf}$ (given by the stripline) is along $y$. Such a configuration excites the uniform Kittel mode of Py ($k^{\rm Py}=0$) throughout $d_1$, and the perpendicular standing spin-wave (PSSW) modes of YIG, (${k}^{\rm YIG} = n\pi/d_2$), due to their magnon-magnon coupling to the uniform Kittel mode of Py \cite{xiong2020probing,xiong2022tunable}.  

In prior experiments, the direct interaction between the MW field ${\rm \bf h}^{\rm rf}$ and the PSSW modes was insignificant, hence often safely neglected \cite{xiong2020probing,xiong2022tunable}. However, depending on the $z$-profile of the MW field (e.g., non-uniform along the YIG thickness direction), the MW ${\rm \bf h}^{\rm rf}$ field can also couple to the PSSW modes directly, contributing to an additional (photon) excitation channel, concurrent with the Py (magnon) channel.  

\subsection{A. Dual-channel interference} 

We first develop an analytical model to predict the coupled dynamics under such a dual-channel excitation. The orientation of the normalized initial equilibrium magnetization at the interface ${\rm \bf m}^{\rm 0, Py (YIG)}(z=0)$ is obtained by minimizing the total magnetic free energy density of the exchange-coupled bilayer. The normalized magnetization of Py and YIG at the interface can be written as: ${\rm \bf m}^{\rm Py (YIG)}(z=0,t)={\rm \bf m}^{\rm 0, Py (YIG)}+\Delta {\rm \bf m}^{\rm Py (YIG)}(z=0,t)$. The second term describes the magnetization variations at the interface due to the dual-channel excitations, which can be written as: 
\begin{equation}
    \begin{aligned}
        \Delta {\rm \bf m}^{\rm Py}(z=0,\omega) = & \boldsymbol{\chi}^{\rm Py}(k^{\rm Py}=0,\omega) \times \\
                                            & [\textrm{\bf h}^\textrm{Py}(\omega) + J_{\rm int} \delta(z) \Delta {\rm \bf m}^{\rm YIG}(z=0, \omega)],
    \end{aligned} 
\label{Eq:eq1}
\end{equation}
\begin{equation}
    \begin{aligned}
        \Delta {\rm \bf m}^{\rm YIG}(z=0,\omega) = & \boldsymbol{\chi}^{\rm YIG}(k^{\rm YIG},\omega) \times \\
                                            & [\textrm{\bf h}^\textrm{YIG}(\omega) + J_{\rm int} \delta(z) \Delta {\rm \bf m}^{\rm Py}(z=0, \omega)],
    \end{aligned} 
\label{Eq:eq2}
\end{equation}
\noindent where $\textrm{\bf h}^\textrm{Py}(\omega)$ and $\textrm{\bf h}^\textrm{YIG}(\omega)$ are the rf magnetic fields in the Py and YIG, respectively (the photon channel), $J_{\rm int} \delta(z) {\rm \bf m}_i^{\rm Py} \sim {\rm \bf H}_i^{\rm exch,YIG}$ is the interlayer exchange field on YIG and $J_{\rm int} \delta(z) {\rm \bf m}_i^{\rm YIG} \sim {\rm \bf H}_i^{\rm exch,Py}$ is the interlayer exchange field on Py (magnon channel), $J_{\rm int} = 2J / \mu_0 (M_s^{\rm Py}+M_s^{\rm YIG})$, and $i = x, y, z$. Here, $\mu_0$ is the vacuum permeability; $M_{\rm s}^{\rm Py}$ and $M_{\rm s}^{\rm YIG}$ are the saturation magnetization of Py and YIG; a positive $J$ indicates a ferromagnetic (FM) type interlayer exchange coupling, while a negative $J$ indicates an antiferromagnetic (AFM) type coupling. \textcolor{black}{$\boldsymbol{\chi}^{\rm Py}(k^{\rm Py}=0,\omega) \approx \frac{{\rm \bf A}^{\rm Py}}{g^{\rm Py}({\rm \bf H}^{\rm bias})}$ and $\boldsymbol{\chi}^{\rm YIG}(k^{\rm YIG},\omega)\approx\frac{{\rm \bf A}^{\rm YIG}}{g^{\rm YIG}({\rm \bf H}^{\rm bias})}$ are the intrinsic susceptibility tensor, which can be computed analytically by solving the linearized LLG equation under weak external stimuli, see Supplemental Materials \cite{SM} (Sec. S1 and S2) for calculation details and the reference therein \cite{song1994giant,hu2016multiferroic,li2020coherent,ding2020sputtering}. The ${\rm \bf A}^{\rm Py(YIG)}$ and $g^{\rm Py(YIG)}({\rm \bf H}^{\rm bias})$ are the amplitude coefficient and Lorenz lineshape of Py and YIG resonances, respectively.}

\begin{figure*}[htb]
 \centering
 \includegraphics[width=7.2 in]{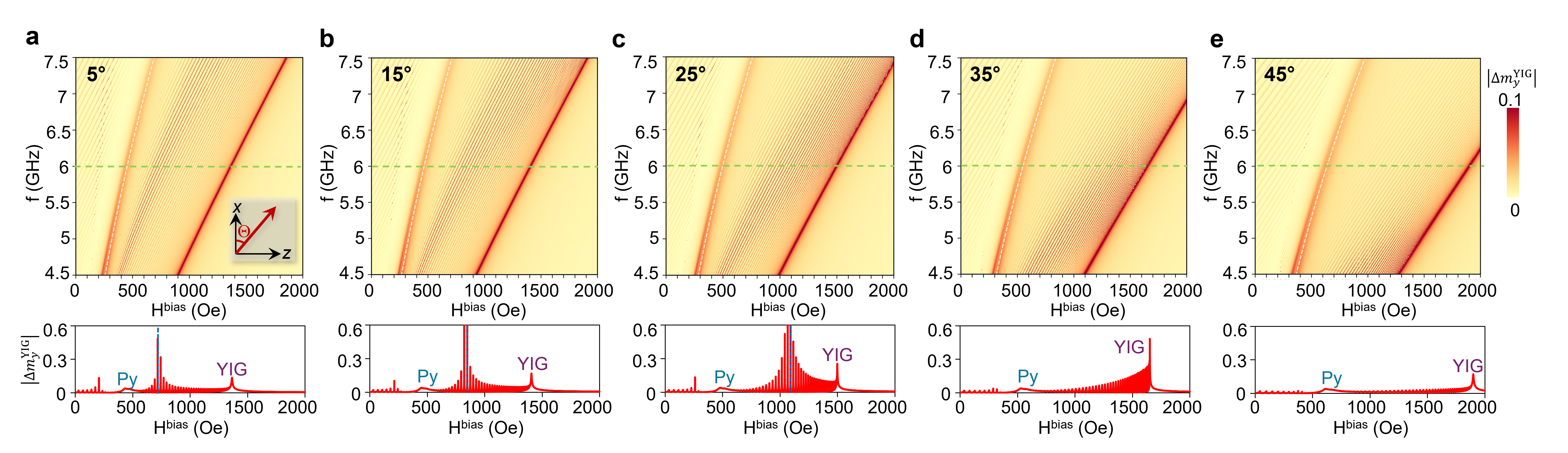}
 \caption{(a-e) The effect from the oblique external bias field ${\rm \bf H}^{\rm bias}$ under a dual-channel excitation and assuming an FM interlayer coupling between YIG and Py. Top panel: 2D $f - H$ dispersion contour plots with selective ${\rm \bf H}^{\rm bias}$ angles $\Theta$, at (a) $5^\circ$, (b) $15^\circ$, (c) $25^\circ$, (d) $35^\circ$, (e) $45^\circ$. Bottom panel: example 1D signal-line trace at 6-GHz for each field angle. }
 \label{fig3}
\end{figure*}

It should be noted that Eqs. (1-2) describe the interfacial magnetization variation resulting from the intrinsic susceptibilities of the Kittel magnon mode in the Py to a specific PSSW mode of YIG ($k^{\rm YIG}$). Since there are often multiple, concurrent PSSW modes, the overall magnetization variation is a superposition of all the excited PSSW modes. Furthermore, because the ultimate excitation source is still the MW drive, whether directly on YIG or indirectly via Py,  the total effective susceptibility of YIG PSSW modes for the dual channel case can be written as:  
\begin{equation}
    \textcolor{black}{|\Delta {\bf m}^{\rm YIG}|= {|\boldsymbol{\chi}}^{\rm dual} \cdot {\rm \bf h}^{\rm rf}|}, 
\label{Eq:eq3}
\end{equation}
\noindent where, 
\begin{equation}
    \begin{aligned}
        \textcolor{black}{{\boldsymbol{\chi}}^{\rm dual} = {\bf K}^{-1}[{\boldsymbol{\chi}}^{\rm YIG}(J_{\rm int}\delta(z){\boldsymbol{\chi}}^{\rm Py}+{\bf I})]}, 
    \end{aligned} 
\label{Eq:eq4}
\end{equation}
\noindent and, ${\bf K} = {\bf I}-[J_{\rm int}\delta(z)]^2 \frac{{\rm \bf A}^{\rm YIG}}{g^{\rm YIG}({\rm \bf H}^{\rm bias})} \frac{{\rm \bf A}^{\rm Py}}{g^{\rm Py}({\rm \bf H}^{\rm bias})}$ is the coupling coefficient matrix and $\rm \bf I$ is the identity matrix. Here, the ${\boldsymbol{\chi}}^{\rm YIG}(J_{\rm int}\delta(z){\boldsymbol{\chi}}^{\rm Py}+{\bf I})$ term in Eq.\ref{Eq:eq4} manifests the combined but interfering contributions due to the magnon (
$ {\bf K}^{-1} \boldsymbol{\chi}^{\rm{YIG}} J_{\rm int}\delta(z) \boldsymbol{\chi}^{\rm Py} $ 
) and photon ( 
$ {\bf K}^{-1} \boldsymbol{\chi}^{\rm{YIG}} $ ) coupling channels, which breaks the symmetry of YIG excitation amplitude on either side of the Py Kittel mode. 
Without such an interference, the excitation is governed solely by the symmetric response, resulting in a symmetric YIG spectrum around the Py resonance \cite{xiong2022tunable}. Including the interference term breaks the symmetry through phase-dependent constructive/destructive couplings. When the photon channel is insignificant as in prior situations \cite{xiong2020probing,xiong2022tunable}, Eq.\ref{Eq:eq4} then regresses to ${\bf K}^{-1}[J_{\rm int}\delta(z){\boldsymbol{\chi}}^{\rm YIG}{\boldsymbol{\chi}}^{\rm Py}]$, and the above symmetry breaking vanishes. See the Supplemental Materials \cite{SM} for the detailed calculations (Sec. S3), discussion of the spectral asymmetry (Sec. S4), and parameter values (Sec. S6), and the references therein \cite{zhang2021strong,xiong2020probing,xiong2022tunable,chikazumi1997physics}.   


\begin{figure*}[htb]
 \centering
 \includegraphics[width=7.2 in]{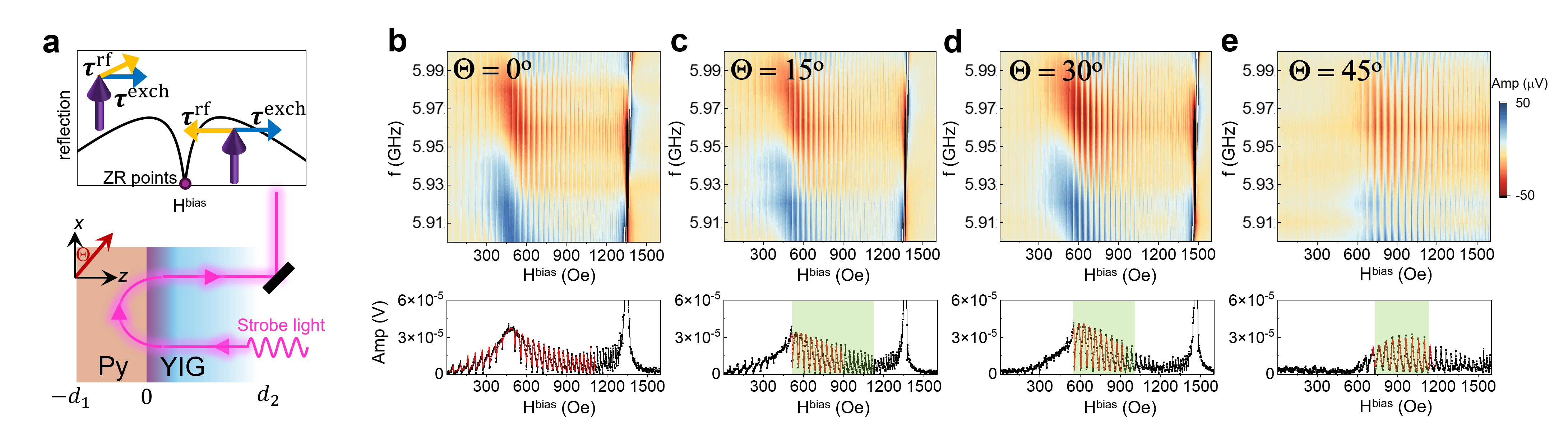}
 \caption{(a) Schematic illustration of the reflective magneto-optic measurement using the strobe light spectroscopy. The ZR point in the magneto-optic spectrum is characterized by a destructive interference between the two involving torques, $\stau^{\rm rf}$ and $\stau^{\rm exch}$, to the magnetization precession. (b-e)  Experimental $f - H$ dispersion contour plots at different ${\rm \bf H}^{\rm bias}$ angles, $\Theta = 0^\circ, 15^\circ, 30^\circ, 45^\circ$. The lower panel shows example 1D signal-line traces (amplitude) at 6-GHz for each angle. \textcolor{black}{The (green) shadowed area highlights the amplified regime.} }
 \label{fig4}
\end{figure*}

\subsection{B. Magnon dispersions} 

We show the calculated $f-H$ dispersions in the cases of YIG excitation via a single channel (magnon only) and a dual channel (magnon and photon) schemes, near the frequency range of 4.5-7.5 GHz encompassing the Py Kittel mode along with a series of YIG's PSSW modes in the vicinity, see Fig.\ref{fig2}. These excited PSSW modes, given the thickness of YIG, reaches high indices up to $n= 57$ with wavelengths down to $\sim 105$ nm. Fig. \ref{fig2}(a,b) compares the single- and dual-channel excitations under an AFM interfacial coupling of Py and YIG. For the single-channel case, Fig.\ref{fig2}(a), the YIG's PSSW spectrum is rather symmetric at either side of the Py Kittel mode. This result is in line with prior experimental reports in which the direct photon channel excitation is omitted \cite{xiong2020probing,xiong2022tunable}. 

However, adding such a photon channel causes prominent asymmetry in the excited PSSW modes along the two sides of the Py Kittel mode: in the dual-channel case, the low-field PSSWs become significantly bolstered in the amplitude, manifesting a large asymmetry with respect to the center Py Kittel resonance, Fig. \ref{fig2}(b). This effect is also seen from the selected single line trace of YIG's magnetization amplitude calculated at 6-GHz. {As discussed above, such an asymmetry is caused by the ${\boldsymbol{\chi}}^{\rm YIG}(J_{\rm int}\delta(z){\boldsymbol{\chi}}^{\rm Py}+{\bf I})$ term in Eq. \ref{Eq:eq4}. Likewise, Fig.\ref{fig2}(c,d) compare the same situation but under an FM interfacial coupling between Py and YIG. Notably, depending on the type of the interface coupling (AFM \cite{qian2024unraveling} or FM \cite{quarterman2022probing}), the excitation asymmetry exhibits opposite trends. For the dual-channel case with FM coupling, the high-field PSSW modes got bolstered, see Fig.\ref{fig2}(d), compared to the symmetric, single-channel scheme, in Fig.\ref{fig2}(c). 

Another notable feature is that some specific PSSW modes have significantly larger amplitudes than the others, which is the case for both low-field and high-field PSSWs. \textcolor{black}{This mode-selective enhancement arises from the structure of the inverse coupling coefficient matrix, expressed as ${\bf K}^{-1} = \text{adj}({\bf K})/\det({\bf K})$. Specifically, at certain ${\rm \bf H}^{\rm bias}$ values where the YIG PSSW modes and Py Kittel mode \textcolor{black}{co-resonate} (i.e., the real part of $\det({\bf K})$ equals zero), each susceptibility peak corresponds to a particular resonant mode. However, among these peaks, the most pronounced one does not occur where the imaginary part of $\det({\bf K})$ is minimized, nor where the adjugate matrix element is maximized. Rather, it arises at a specific wavenumber $k$ where the ratio $\left|\text{adj}({\bf K})/\det({\bf K})\right|$ is maximized, reflecting an optimal balance between the spectral dissipation and the effective mode-to-mode coupling strength within the hybrid system.} In addition, the mode-dependent dissipation rate further influences the amplitude of each individual PSSW mode. Detailed analyses are provided in the Supplemental Materials (Sec. S5) ~\cite{SM}.

\subsection{C. Oblique-field dependence} 

Our model also predicts that such selective amplification or attenuation has a sensitive response to an oblique bias field, ${\rm \bf H}^{\rm bias}(\Theta)$. As an example, we show in Fig.\ref{fig3}, the calculated $f - H$ dispersions (top panel) of the bilayer (assuming FM coupled) under a dual-channel excitation, and the selective 1D traces at 6-GHz (bottom panel). Three notable spectral characteristics can be predicted. First and foremost, the asymmetric amplification of the PSSW modes becomes more pronounced as the angle $\Theta$ increases. Such field-controllable asymmetry is attributed to the effect of a tilted initial equilibrium magnetization ${\rm \bf m}^{\rm 0}$ on the ${\boldsymbol{\chi}}^{\rm YIG}(J_{\rm int}\delta(z){\boldsymbol{\chi}}^{\rm Py}+{\bf I})$ term} in Eq.\ref{Eq:eq4}. Second, as the ${\rm \bf H}^{\rm bias}$ rotates toward the z axis, the packet of PSSW modes receiving more significant amplification shifts significantly to the high field (lower-index modes) until the mode number reaches zero (i.e., the Kittel mode of YIG). This mode-selective amplification is due to the influence of the titled ${\rm \bf m}^{\rm 0}$ on the coupling coefficient matrix ${\bf K}$, as can be seen from the corresponding curves of ${\bf K}({\rm \bf H}^{\rm bias})$ in Supplemental Materials (Sec. S5) \cite{SM}. Third, the Py and YIG Kittel modes both shift to higher ${\rm \bf H}^{\rm bias}$ merely due to the obliqueness of the bias field, and this shift is particularly significant for YIG as indicated by the vertical dashed lines in the example 1D line traces.

\section{III. Experiments}

Next, we show our experiment that confirms the analytical model presented herein. The calculated magnetization variations in Eq.\ref{Eq:eq3} and \ref{Eq:eq4} can be experimentally detected via dynamic magneto-optic Kerr and Faraday effects. We employed an infrared-band strobe light probe \cite{xiong2024phase} and detected the coupled dynamics in the reflective geometry, see Fig.\ref{fig4}(a). We use a coplanar waveguide (CPW) with a very thin signal line (250 $\mu$m), compared to the lateral size of the YIG/Py bilayer \textcolor{black}{($3 \times 3$ mm$^2$)} so as to amplify the excitation by the non-uniform MW ${\rm \bf h}_{\rm rf}$ field (photon channel) in parallel to the Py exchange field (magnon channel). Due to the relatively thick YIG layer, the magnon-magnon coupling between Py Kittel and YIG PSSWs falls into the MIT coupling regime \cite{xiong2020probing,xiong2022tunable}, where the reflective magneto-optic spectra exhibit sharp dips analogous to the ZR phenomena \cite{zhang2014coherent,yoo2009chip,soukoulis2011past}, i.e., spectrum locations where the reflection coefficient is locally tuned to zero, see Fig. \ref{fig4}(a).

We show in Fig. \ref{fig4}(b-e) the measured magneto-optic ZR spectra obtained at different ${\rm \bf H}^{\rm bias}$ orientations, $\Theta = 0^\circ, 15^\circ, 30^\circ, 45^\circ$. The 2D contour plots are generated from the individual Re[$V(f,H)$] spectra (amplitude and phase) acquired by sweeping the ${\rm \bf H}^{\rm bias}$ at a given frequency (from 5.9 to 6.0 GHz). An example 1D signal trace at 6.0 GHz (amplitude) is shown in the lower panel of Fig.\ref{fig4}(b-e). First, compared to prior reports \cite{xiong2020probing,xiong2022tunable}, the direct excitation by the photon channel is effective in the present experiment. This is evidenced from the prominent excitation of a comprehensive series of PSSW modes (reaching $n > 40$) at a wide range of ${\rm \bf H}^{\rm bias}$, even away from the Py Kittel mode profile, see Fig.\ref{fig4}(b), manifesting a series of sharp and closely-packed ZR points throughout the measured ${\rm \bf H}^{\rm bias}$ range. 

Notably, the excitation amplitude is asymmetric, i.e., more pronounced on the right side of the Py Kittel resonance (high-${\rm \bf H}^{\rm bias}$) but less on the left side (low-${\rm \bf H}^{\rm bias}$). This is consistent with the prediction of our model assuming an FM interface coupling between the Py and YIG. Such an amplitude asymmetry becomes more pronounced as the angle of ${\rm \bf H}^{\rm bias}$ increases. At $\Theta=15^\circ$, the high-${\rm \bf H}^{\rm bias}$ excitation regime exceeds that for the low-${\rm \bf H}^{\rm bias}$ one, resulting in a largely suppressed ZR spectrum at low ${\rm \bf H}^{\rm bias}$. At $\Theta=30^\circ$, the amplitude of the high-${\rm \bf H}^{\rm bias}$ regime continues to grow, while the low-${\rm \bf H}^{\rm bias}$ counterpart is almost completely suppressed, showing nearly only the Py profile alone. At $\Theta=45^\circ$, \textcolor{black}{despite the overall suppressed resonance profile of Py (due to the oblique ${\rm \bf H}^{\rm bias}$), the excitation asymmetry continues to grow, in which the signal amplitude at the low-${\rm \bf H}^{\rm bias}$ regime is almost entirely nullified}, effectively changing the collection of distinct ZR modes to an extended ZR band. 

In conjunction with the enhanced excitation amplitude, the field window for observing such an enhancement narrows as the angle $\Theta$ increases -- one of the main features that our analytical model also predicts, see Fig.\ref{fig3}: as the $\Theta$ angle increases, the energy appears to be more concentrated in just a few modes near the Py Kittel mode rather than evenly spreading out. Experimentally, at $\Theta=0^\circ$ and $15^\circ$, the enhanced excitation extends all the way up to the long-wavelength PSSW modes, i.e., those near the YIG Kittel mode at higher ${\rm \bf H}^{\rm bias}$. While at $\Theta=30^\circ$ and $45^\circ$, such a window narrows only up to the vicinity of the Py Kittel mode. 

\section{IV. Summary and Outlook} 
 
In summary, we theoretically and experimentally demonstrated a tunable magneto-optic ZR spectrum in a magnon-magnon coupled, YIG/Py bilayer subject to a dual-channel excitation scheme. \textcolor{black}{While the excitation efficiency of each channel is vital in its own right, we show that their combination augments a new leverage to effectively modulate the coupled dynamics. Namely,} the co-operational Zeeman and exchange torques can lead to destructive or constructive interference, and result in suppressed or amplified effective excitations of the finite-wavelength modes of YIG. Notably, such an effect is stipulated by the type of interfacial couplings (either AFM or FM), and can be effectively tuned by the oblique external magnetic fields. 

\begin{figure}[htb]
 \centering
 \includegraphics[width=3.5 in]{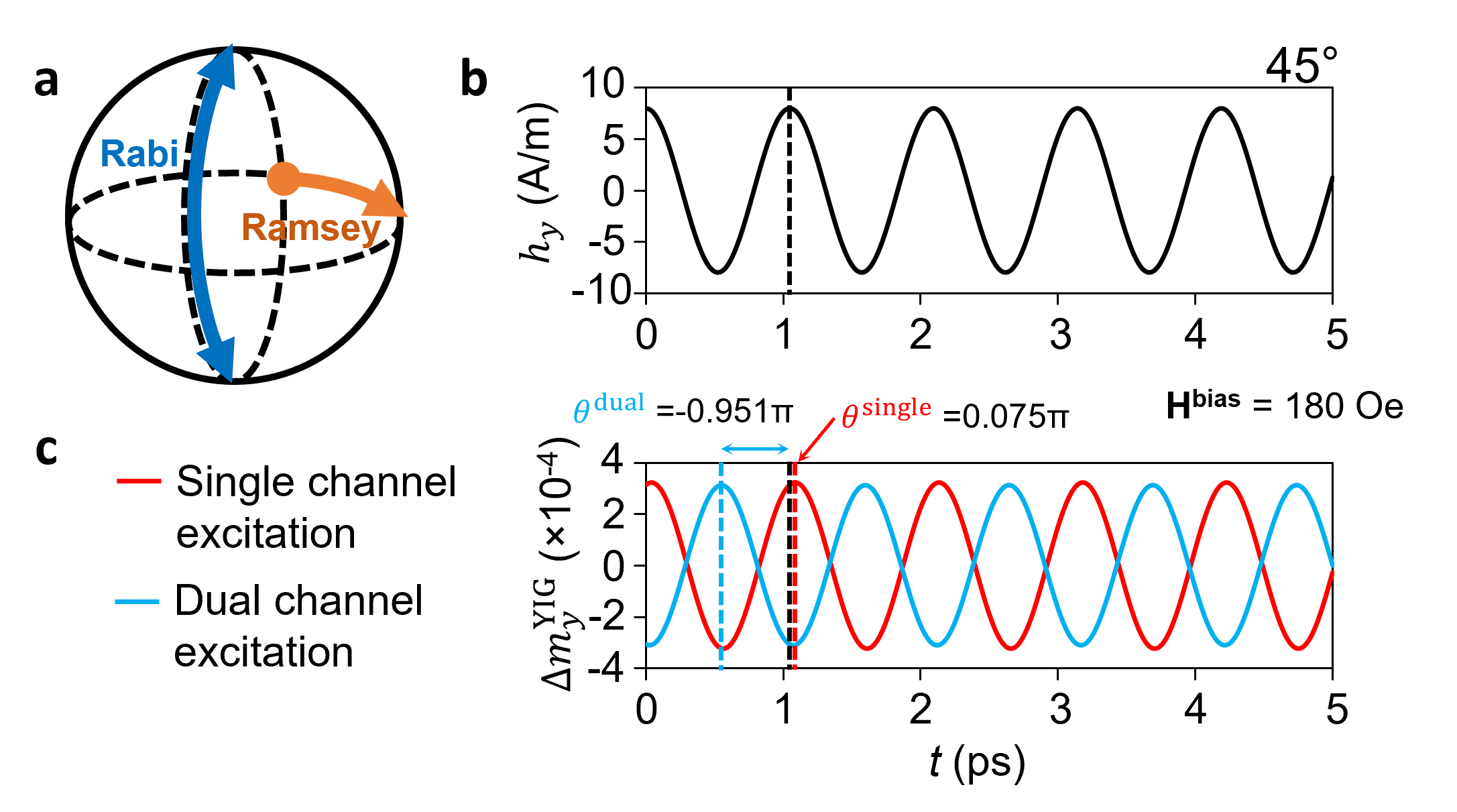}
 \caption{\textcolor{black}{(a) The Bloch sphere of a two-level system representing the magnon-magnon coupling (e.g. a YIG/Py bilayer). The state transfer is either direct $\bf \ket{0} \rightarrow \bf \ket{1}$ (Rabi), or indirect $\bf \ket{0} \rightarrow \bf \ket{\Phi} \rightarrow \bf \ket{1}$ (Ramsey). In the latter, the magnon phase $\Phi$ is used as a controlled state variable for mediating the transfer process. (b) The external MW excitation field, $\rm h_{rf}$, and (c) the temporal evolution of the YIG's $\Delta \textrm{m}^\textrm{YIG}(y)$ precession for the single-channel (red) and dual-channel (blue) cases, in reference to the external $\rm h_{rf}$. The phase shifts are 0.075$\pi$ for the single-channel case, and $-0.951\pi$ for the dual-channel case. }}
 \label{fig5}
\end{figure}

\textcolor{black}{As an application note, the dual-channel mechanism represents a phase coherent operation on the short wavelength modes, which can potentially be exploited to realize accurate phase retarding and switching functionalities in $\bf k$-specific magnon cavitronics. To date, research on magnon cavitronics has been predominantly focused on the Kittel magnon mode  ($\bf k$ = 0) from a single magnetic material, and only in the context of a Rabi process, see Fig. \ref{fig5}(a), which shuffles the two-level states, $\bf \ket{0}$ and $\bf \ket{1}$ bilaterally, within each one’s interaction zone. In this case, the spin precessional phase is excluded from being a state variable, despite that the phase control is highly beneficial to the technological important Ramsey process, see Fig. \ref{fig5}(a), where additional control arises from the equatorial, free-precession regime. For example, using practical parameters computed from this work, we select a scenario where the bias magnetic field is $\bf H^{bias}$=180 Oe and at $\Theta=45^\circ$. We focus on the high-$\bf k$ mode (specifically, near the $n=53$ PSSW mode in YIG) at 6 GHz under a reference MW field ${\rm \bf h}_{\rm rf}$, see Fig. \ref{fig5}(b). Figure \ref{fig5}(c) compares the temporal evolution of the YIG's $\Delta \textrm{m}^\textrm{YIG}(y)$ precession for the single-channel (red) and dual-channel (blue) cases. The dual-channel excitation significantly modifies the phase of YIG relative to the driving field as compared to the single-channel case, enabling a pronounced and accurate phase engineering scheme. The phase shifts are 0.075$\pi$ for the single-channel case, and $-0.951\pi$ for the dual-channel case. Such a controllable phase tuning for a specific-$\bf k$ magnon mode can be directly applied in device designs where phase-matched or phase-inverted magnetization responses are required, for instance, spin-wave frequency diplexers~\cite{chuang2011microstrip}, phase-coherent majority gates~\cite{talmelli2020reconfigurable}, and zero-reflection couplers~\cite{wang2018reconfigurable,odintsov2025bilayer}. } 

\textcolor{black}{In particular, while most current efforts have focused on engineering spin-wave circuitry for interfering with state variables and transfer functions. Our results represent a material science route to access the Ramsey cycle in magnon cavitronics leveraging magnetic interactions.} Given the ubiquitousness of both Zeeman and exchange torques in such and similar coupled magnets, our conclusions can be extended to many materials systems alike, including both metallic \cite{qin2018exchange,qin2021nanoscale,liensberger2019exchange,xiong2024phase} and insulator ones \cite{liu2024strong,li2024reconfigurable,fan2025dynamically}. The results can also find usefulness for understanding and manipulating hybrid magnonics spectra under various control knobs, leveraging the combination and interference between different coupling channels, such as harnessing the different interlayer spin-spin coupling schemes (magnon-magnon) and/or optimizing the microwave photon field landscape via tailored waveguide and resonator designs (photon-magnon) \cite{xiong2024combinatorial,xiong2024hybrid}.

\section{Acknowledgments}

The experimental work at UNC-CH was supported by US National Science Foundation (NSF) under award Nos. DMR-2509513 and ECCS-2426642. Y.Z. and J.-M.H. acknowledge the support by the U.S. Department of Energy, Office of Science, Basic Energy Sciences, under Award Number DE-SC0020145 as part of the Computational Materials Sciences Program. The computation in this work used the Bridges system at the Pittsburgh Supercomputing Center through allocation TG-DMR180076 from the Advanced Cyberinfrastructure Coordination Ecosystem: Services \& Support (ACCESS) program, which is supported by NSF Grants No. 2138259, No. 2138286, No. 2138307, No. 2137603, and No. 2138296.

\bibliography{sample}

\end{document}